# An Economical and Efficient Helium Recovery System for Vibration-Sensitive Applications

Zhiyuan Yin,[1 a)] Liya Bi,[1,2 a)] Yueqing Shi,[1] and Shaowei Li [1,2,b)]

[1]*Department of Chemistry and Biochemistry, University of California, San Diego, La Jolla, CA 92093-0309, USA*

[2]*Program in Materials Science and Engineering, University of California, San Diego, La Jolla, CA 92093-0418, USA*

a) These two authors contributed equally to this work.
b) Author to whom correspondence should be addressed: shaoweili@ucsd.edu

We present the design of a helium liquefaction system tailored to efficiently recover helium vapor from either an individual or a small cluster of vibration-sensitive cryogenic instruments. This design prioritizes a compact footprint, mitigating potential contamination sources such as gas bags and oil-lubricated compressors while maximizing the recovery rate by capturing both the boil-offs during normal operation and refilling process of the bath cryostat. We demonstrated its performance by applying it to a commercial low-temperature scanning probe microscope. It features a > 94% recovery rate and induces negligible vibrational noise to the microscope. Due to its adaptability, affordability, compact size, and suitability for homemade setups, we foresee that our design can be utilized across a wide range of experimental measurements where liquid helium is used as the cryogen.

## I. INTRODUCTION

Given the low density and chemical inertness[1], helium (He) is pivotal in various medical[2, 3], industrial[4, 5], and aerospace applications[6, 7]. It provides propulsion and cooling in aerospace contexts[8, 9], facilitates deep-sea diving via mixed gas formulations[10, 11], and is integral to industrial processes like leak detection[12, 13] and semiconductor manufacturing[14, 15]. Its inertness helps create a chemically stable environment, which is critical in applications like zirconium[16] or silicon production[14, 15] and gas chromatography[17, 18]. Its notable thermodynamic properties[1], especially the ability to approach near absolute zero temperatures either as the working gas[19-21] or cryogen[22], make it essential in cryogenic applications such as cooling superconducting magnets[23] and maintaining ultra-low operational temperatures in particle colliders[24], spectrometers[25], microscopes[26, 27], Mossbauer experiments[28], and quantum ion traps[29].

The looming supply-demand imbalance of He[30-32] calls for efficient recovery technologies[24, 33, 34]. Recent data show a 7.7% in production decrease, confronting an estimated 9.3% demand increase in the United States[35]. Furthermore, He escapes into space due to its light weight, making it impossible to recapture once released into the air[1]. Thus, He recycling emerges not simply as a strategy to navigate the challenges of its scarcity but as a crucial approach, aligning with global sustainability and economic considerations and ensuring a stable supply amidst soaring market demand[30, 32, 36, 37].

Existing He recovery schemes for the research labs, mainly gasbag-style systems[33, 34] and closed-cycle designs[28, 29, 38-40], suffer from various drawbacks and very rarely meet both the need and financial situation of individual labs that run vibration-sensitive cryogenic instruments. On the one hand, while mature products exist[33, 34], the gasbag-style recovery systems occupy substantial lab space that is not easily accessible to many research labs. Besides, the gas bag or balloon for temporarily storing the He boil-off together with the oil-lubricated compressor used to further compress He into the medium-pressure storage tanks introduces significant contamination and leakage[33, 34], necessitating thorough purification processes. Normally, the gasbag-style systems take collaborative resources from multiple labs and are usually designed as shared facilities, making them economically unfavorable and even impractical to individual labs. On the other hand, closed-cycle[28, 29, 38-40] designs offer another way to recover He vapor and avoid liquid He (LHe) replenishment. In these schemes, the cryocooler is usually placed in proximity to the cryogenic apparatuses to ensure sufficient cooling power. As a result, the moving parts[41] of the cryocooler very often transmit significant vibrations to the equipment, making these designs challenging for low-vibration applications. While solutions with vibrational decoupling mechanisms[39, 42] have been proposed to mitigate the mechanical noise, they typically require sophisticated engineering and special modifications to accommodate vibration-sensitive instruments[29, 41, 43]. Therefore, it is either economically disadvantaged (as with gasbag-style system and closed-cycle design) or technically difficult (as with closed-cycle design) to implement the existing He recycling solutions to individual or a small cluster of cryostats that are sensitive to vibrations.

In response, our team has developed a simple and highly effective He recovery system that seamlessly integrates with a conventional bath cryostat (CRYOVAC) for the low-temperature scanning probe microscope (LT-SPM) (CreaTec). This setup is designed to optimize cost-effectiveness and operational efficiency. It demonstrates an average daily LHe waste of less than 0.27 liters in a one-year duration, which translates to only 98.5 liters loss after consuming 1,825 liters, including that from all major sources such as daily running, liquid transfer (cryostat refilling and initial cooldown) and servicing the cold head. Our design largely utilizes and repurposes commercially available components, avoiding complications from custom design or development. This solution is shown to have a negligible impact on the operation and performance of the SPM and can easily be adapted for other vibration-sensitive apparatuses.

## II. SYSTEM OVERVIEW

Our strategy involves the direct re-liquefaction of daily He boil-off or vapor from both the SPM bath cryostat and storage dewars (CryoFab CMSH LH100), alongside capturing and storing any excess boil-off during liquid transfer into the medium-pressure tanks. Accordingly, the system mainly comprises two sections: He condensation and compression/storage (**Figure 1** and **Figure 2a-c**), which neatly fit into a small area of less than 4 $m^2$. The condensation section, equipped with a commercial He liquefier cooled by a two-stage pulse tube cryocooler (Cryomech HeRL10 with the PT410 cold head), is rated to liquefy/recondense more than 10/18 L LHe per day. The liquefier terminates with a liquid return line, which is directly inserted into a LHe dewar (receiving dewar, **Figure 1** and **Figure 2b, c**) via its refilling port. This setup allows for immediately (re)condensing the He boil-off into the receiving dewar. When close to full, the receiving dewar can be swapped with another container, the refilling dewar (**Figure 1** and **Figure 2d**), which replenishes the SPM cryostat (**Figure 1** and **Figure 2e, f**). In other words, LHe transfer from the receiving dewar to the refilling dewar is avoided, which usually induces extra He loss during evacuating and cooling the transfer line in the gasbag-style recovery systems. Normally, the liquefier can capture the He boil-off from the cryostat together with the receiving and refilling dewars. However, when initially cooling down or refilling the cryostat, the momentary boil-off rate will surpass the maximum liquefaction capacity. To maximize the recycling efficiency, the compression/storage section, mainly comprising a modified oil-free air compressor (California Air Tools 600040CAD) and 130-gallon gas storage tanks, is used to temporarily store this excess He boil-off and can hold up to 4,677 L He gas (6.24 L LHe) at 125 psi (**Figure 1**, **Figure 2a, b, Appendix A4** and **Appendix B**). A typical replenishment for our SPM cryostat adds less than 2,100 L He gas to the He boil-off storage tanks, which will be liquefied into the receiving dewar before the next refill (usually every 3 days). By recondensing the He boil-off from both daily machine running and cryostat initial cooldown/refilling, we manage to recover over 94% of the LHe our lab consumes.

The successful implementation of our recycling system on the SPM relies on the ultra-high purity of He boil-off, the long-term stable inline He vapor pressure, and the highly effective vibration reduction. To ensure the inline He purity, we mainly utilized stainless-steel (SSL) tubing with ConFlat (CF), vacuum coupling radiation (VCR), or Swagelok termination for the He vapor lines (**Figure 1** and **Figure 2**), with only few exceptions where Klein Flansche (KF), e.g., the outlet of SPM cryostat and the connection to air compressor, or national pipe tapered (NPT) connector, e.g., the safety valves and the connection to He boil-off storage tanks (**Appendix A4**), is adopted. Besides, the attachment of the LHe dewars to the He vapor lines is made contamination-free via the self-lock Swagelok quick connectors (**Appendix A1**). Consequently, the daily He boil-off from the SPM cryostat and two LHe dewars, accounting for more than 80% of the He recycled, is well isolated from the lab environment and doesn't need further purification before recondensation. Nevertheless, we found that the excess

He boil-off that is compressed and stored in the He boil-off storage tanks during liquid transfer can clog the cold head, potentially due to a low amount of contamination induced by the air compressor. To improve the purity of this portion of He boil-off, we applied He-compatible sealants (Gasoila GE16) and Teflon tapes to the NPT connections on the air compressor motor and the storage tanks and remove the contaminants (primarily water, oil, and carbon dioxide as discussed in **Section IV**) in the compressed He with a purifier (**Figure 1** and **Figure 2a, b**) before feeding this He to the liquefier to condense. It is worth noting that this purifier is simply built by immersing an SSL coil into the liquid nitrogen ($LN_2$) held in a commercial cryogenic freezer (MVE XC47/11-6) (**Appendix C3**). In this way, the He flow through the SSL coil is very effectively purified and won't compromise the cold head performance for an extended time.

Both the stable He vapor pressure and the low vibration induction are critical to the performance of LT-SPM because the former elongates the low-temperature holding time and keeps the sample clean[27] by maintaining an undisturbed scanner temperature while the latter guarantees a lower noise level for high-quality data acquisition[44]. However, various operations can increase the daily boil-off rate of the SPM cryostat, such as transferring the tip and sample, and result in an elevated He vapor pressure. Accordingly, we took three measures to deal with the fluctuating He pressure without interfering with the running of SPM. Firstly, the cold head compressor (**Figure 1** and **Figure 2b, c**), customized with an inverter, can increase/decrease the compressing frequency or liquefaction power to match a higher/lower He vapor load, thus targeting the He pressure (monitored by a pressure transducer) for a preset value. Secondly, when the total He boil-off rate is smaller than the minimal liquefaction rate, a heater inside the cryocooler supplies extra He vapor load by heating up the condensed He in the liquefier and hence always keeps the inline He pressure (monitored by a pressure transducer) around the setpoint. In addition, we wrote a LabVIEW program (**Appendix B**) to monitor the He pressure in both the vapor line and the He boil-off storage tanks via digital pressure gauges (Additel 681) (**Figure 1** and **Figure 2b**) and control a mass flow controller (MKS MFC GE50A) located in between the purifier and He vapor line (**Figure 1 and Appendix A3**). This program automatically regulates the purified He flow into the vapor line through the flow controller according to the He vapor pressure and will cut off the flow when the storage tank pressure is below a setpoint, e.g., 20 psi, to prevent contamination. As a result, the He vapor pressure is maintained to be close to the setpoint pressure with a $\pm$ 0.01 psi variation most of the time. To reduce the vibration transmission from the recycling system to the SPM, we adopted three engineering designs. First of all, a pulse tube instead of a Gifford-McMahon cryocooler was used because the former[19] generates smaller vibration than the latter[20] without the need for a moving piston in the cold head. Besides, the cryocooler and other vibration-introducing instruments, including the air compressor, are placed in a remote room from the SPM (**Figure 1** and **Figure 2a-c**). Finally, the SPM cryostat is connected to the He vapor line via an SSL bellow to dampen any residual mechanical vibration from the recycling system further. With all these measures taken, the SPM can run uninterruptedly except during the refilling of the cryostat or servicing of the cryocooler, as mentioned below.

## III. SYSTEM OPERATION

### A. Initial cooldown of the liquefier cold head

To get started, we purge the whole He vapor line multiple times and fill it with ultra-high purity (UHP) He gas. The receiving dewar is sequentially precooled to LHe temperature with the refilling dewar via direct liquid transfer since our cold head is not powerful enough to cool down directly at either room temperature or 77 K dewar. The He boil-off from cooling down the receiving dewar can be compressed into the He boil-off storage tanks to increase recycling efficiency by attaching the dewar vent port to the vapor line via self-lock Swagelok quick connectors (**Appendix A1 and C1**). Afterward, the liquefier return line is inserted into the cold receiving dewar while being purged with the UHP He. A UHP He cylinder (**Figure 1** and **Figure 2b, c**) is used to supply the He for cooling down the cold head, which normally takes ~2 hours to reach 4.2 K under the maximum power (**Figure 3a**).

### B. Initial cooldown of the SPM cryostat

Once the cold head is functioning, we cool down and fill the SPM cryostat with LHe. Before filling it with LHe, the SPM cryostat is vented to the atmosphere through a bypassing valve (Appendix A2 and C2). Once the nitrogen in the cryostat is evacuated, we transfer LHe from the refilling dewar to it and connect it to the He vapor line to compress the excess boil-off into the He boil-off storage tanks. Generally, this process produces more He gas than the capacity of our 130-gallon tanks. Thus, we need to vent some He boil-off, but the portion is small according to our overall >94 % recycling efficiency. This issue can be easily resolved by adding an extension storage tank (**Appendix A4**). When the cryostat is full, we isolate it from the vapor line and vent it to the air again before removing the LHe transfer line to avoid venting the whole recycling line. After the cryostat is capped, we reconnect it to the vapor line for recycling. The refilling dewar is also attached for recovery afterward.

### C. Routine maintenance

#### *1. Refilling the SPM cryostat*

Refilling the SPM cryostat is similar to initially cooling it down, but much less He boil-off is generated. Most of this gas can be compressed into the storage tanks except the part used for evacuating and cooling the liquid transfer line. Besides, we have to isolate and vent the cryostat right before refilling to keep the rest sections of the recycling system unaffected (**Appendix A2 and C2**). It then follows the same procedures as those of the initial cooldown.

### 2. Regenerating the purifier

The 3/8” O.D. SSL coil immersed in $LN_2$ in the purifier can be clogged by frozen contaminants over time. Hence, regeneration is necessary to resume the capability of the coil. Normally, the coil is blocked twice to three times a week and can be unclogged by lifting the top two loops up from the liquid level and blowing them with a hair dryer for several seconds since other gases like water and carbon dioxide usually freeze close to the liquid surface. Any potential contaminants escaping from the coil and heading towards the flow controller during this process will be purged via a Swagelok bellow valve (**Appendix A3**) before redirecting the purified He into the recycling line (**Appendix C3**). This method doesn’t work as effectively after a month because the contaminants are eventually pushed to the lower loops of the coil. In that scenario, we replace the dirty coil with a clean one. This is made reproducible by using Swagelok connections.

### 3. Swapping the receiving dewar and refilling dewar

When the receiving dewar is close to full, we swap it with the nearly empty refilling dewar with minimal interruption to the liquefier (**Figure 3c**). This is done by first isolating the liquefier from all tanks and then blowing the return line outlet with a heat gun while it’s being purged with UHP He to avoid icing, and finally inserting the returning line into the empty dewar (**Appendix C4**). We have found it the most efficient to keep the liquefier running at its lowest power during this process.

### 4. Decontaminating the liquefier cold head

The performance of our cold head is sometimes compromised by contaminants which we suspect to be mostly nitrogen since our purifier doesn’t effectively trap it. At this stage, we isolate the LHe dewars and SPM cryostat from the liquefier, warm the cold head up to 80 K, then cool it down and reconnect all tanks (**Appendix C5**). This proves to very efficiently remove the contaminants in most cases, usually once every several months. We also found it desirable to warm the cold head up to room temperature once a year to fully clean it (**Appendix C5**). Decontaminating the cold head usually takes less than 6 hours if only warming up to 80 K and can take nearly one day if warming up to room temperature (**Figure 3a, b**). This operation can be integrated with the dewar swapping introduced previously to reduce the re-liquefication downtime further.

### 5. Replenishing LHe

Our setup has slight He loss during the initial cooldown, refilling, purging, and decontamination. Hence, LHe replenishment is necessary. This can be achieved by either providing gas to the recovery line using compressed gas cylinders or attaching the vent port of a commercial LHe dewar (LHe replenishing dewar in **Figure 1**) to the He vapor line. The former is more convenient for our purposes because of the low loss rate (**Appendix C6**). While utilizing UHP-grade gas is preferred, we've opted to use industrial-grade He cylinders to reduce costs. The impurities in the industrial-grade He do not obviously compromise the performance of the liquefier, provided they are effectively removed through the purifier.

## IV. PERFORMANCE

### A. Purity of the He in the boil-off storage tanks

The oil-free air compressor we used for compressing the excess He boil-off is low-cost and easy to maintain. However, it is designed for compressing air for pneumatic tools and isn't optimized for He compression. Therefore, it unavoidably introduces contaminants to the He gas. To mitigate He leakage and prevent air contamination, we have applied polytetrafluoroethylene seals to all connecting joints of the compressor and storage tanks. **Figure 4** depicts the mass spectrum of the He gas compressed into the storage tanks, indicating a purity level above 99.5%. The main impurity detected is hydrogen (0.4%), likely permeating through the SSL tubing in the recovery line[45]. As hydrogen solidifies around 14 K, we anticipate its accumulation close to the second stage of our pulse tube cryocooler, which is evident from the warming-up curve of the cold head (**Figure 3b**). However, no other adverse effects due to hydrogen contamination have been observed in our daily operation. The remaining impurities (<0.1%) consist of a mixture of water, nitrogen, oxygen, carbon dioxide, and a trace amount of organic compounds estimated to originate from the compressor lubricant. The majority of these impurities have condensation temperatures above 80 K and can be effectively removed with our cold-trap purifier. A low density of impurities, primarily nitrogen and oxygen, will need to be removed by warming up the cold head on a regular basis, as introduced previously.

### B. Liquefier performance

The liquefier can easily capture our 5 L daily LHe consumption. To optimize efficiency, we intentionally adjust the cold head compressor frequency to its minimum level using the inverter. Operations like reading cryostat He level through the superconducting level meter or loading a new tip/sample from room temperature would increase the boil-off rate, causing a transient pressure change. As mentioned above, the inverter, together with the mass flow controller, is programmed to respond quickly to this pressure change. Consequently, both the cold head temperature and the inline He pressure are highly stable (**Figure 5a, b**). The average daily loss, accounting for all sources, is less than 0.27 L out of 5 L liquid consumption, where most of the loss is from the refilling and unclogging of the system. This loss rate can be further reduced by optimizing the refilling and decontamination procedures.

### C. SPM performance

The SPM we use can operate as either a scanning tunneling microscope (STM) or an atomic force microscope (AFM). Its junction temperature remains stable during prolonged normal operation (**Figure 5c**) due to the stable He vapor pressure. To evaluate the system's noise level, we focused on the STM mode and analyzed the tunneling current with and without being connected to the recycling system (**Figure 6**). Upon connecting to the reliquefying system, the measured noise level remained comparable with measurements taken without connection. The fast Fourier transform (FFT) analysis of the tunneling current revealed several noise peaks, including 60 Hz and its overtones from AC power. Only a few additional noise peaks related to the He liquefier can be identified.

With no obvious interruption from the liquefier, the data quality of the STM is unaffected. In the STM community, detecting single molecule vibrations using inelastic electron tunneling spectroscopy (IETS)[46, 47] is commonly regarded as one of the measurements requiring extremely high system stability. Here, we utilize the IETS of a surface-adsorbed CO molecule and topographic measurements of other materials as test systems to calibrate the performance of the SPM with the attached He recycling system. Atomic resolution and IETS measurements both remain unaffected in our setup. **Figure 7** displays an image of the Cu(100) single crystal surface (**Figure 7a**), the NaCl thin film grown on Cu(100) (**Figure 7b**), and two CO molecules adsorbed on the Cu(100) surface (**Figure 7c**) when the He recycling system is active, along with the IETS spectrum of individual CO molecule (**Figure 7d**). The Cu and NaCl lattice structures are clearly shown under sub-nano resolution, and the hindered transitional (HT) and rotational (HR) modes of the CO are resolved by the STM-IETS[48].

## V. CONCLUSION

In summary, we have successfully designed and implemented an efficient and stable He recovery system, demonstrating its performance with a commercial LT-SPM. This setup exhibits an impressive recycling rate, maintaining a minimal He loss rate of less than 0.27 L/day out of 5 L/day usage. A notable feature is its low vibration level, which, as confirmed by noise analysis and SPM performance, remains consistent regardless of the recovery system's operational status. Our system primarily leverages commercially available components, such as the gas compressor commonly used in the construction and automotive industries, to minimize fabrication costs. This design can easily be adapted for other experimental measurements with a comparable daily consumption rate and is a cost-efficient solution for individual research groups that usually can't afford a large facility.

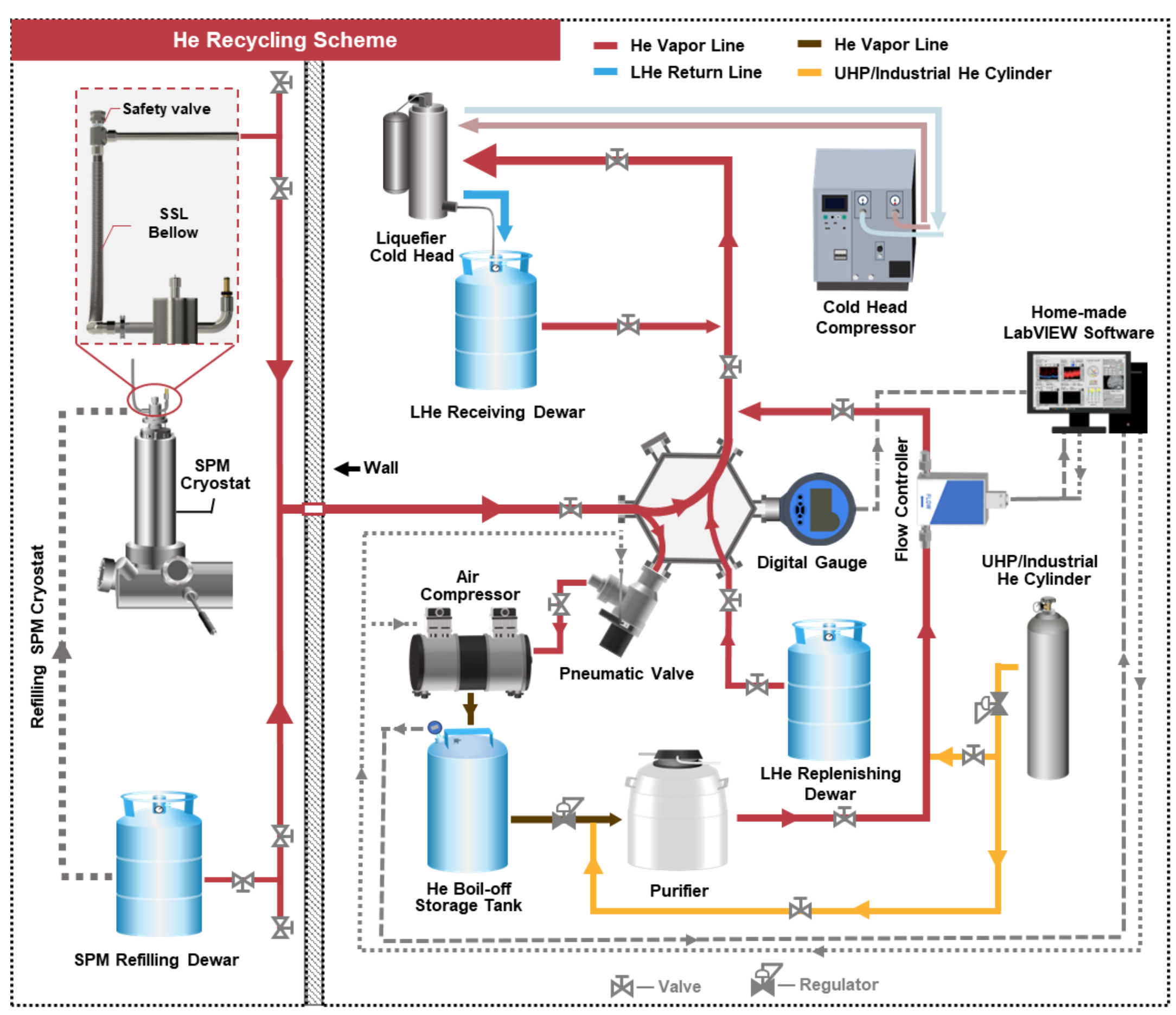

FIG. 1 Schematic layout of the He recovery system.

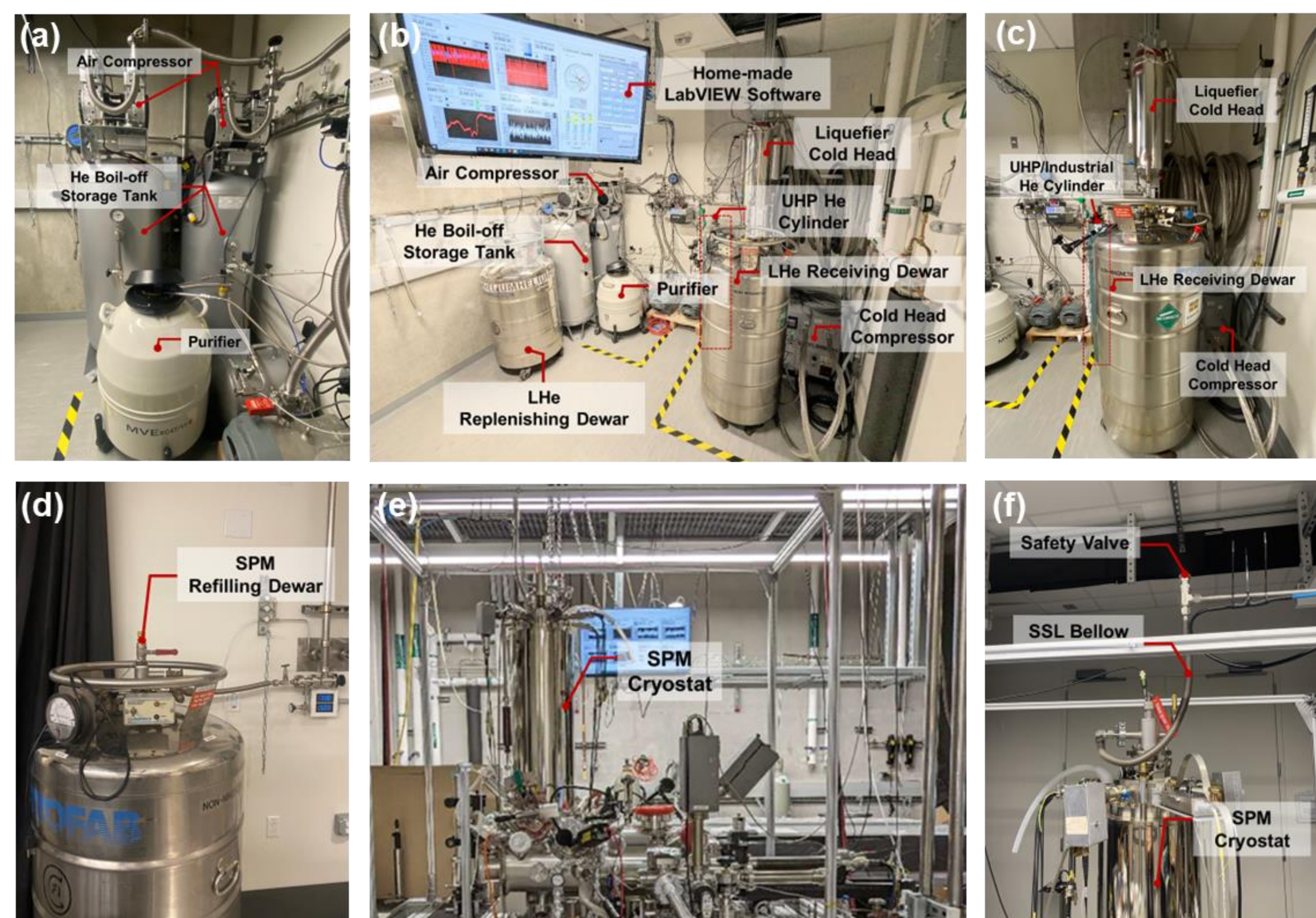

FIG. 2 Photos of the He recycling system. (a) Picture of the air compressor, the He boil-off storage tanks and the purifier. (b) Overview of the He recycling room. (c) Picture of the liquefier, the LHe receiving dewar, and the cold head compressor. (d) View of the parking station of the refilling dewar. (e) Overview of He vapor line at the SPM side. (f) Zoom-in image of the connection to the SPM cryostat.

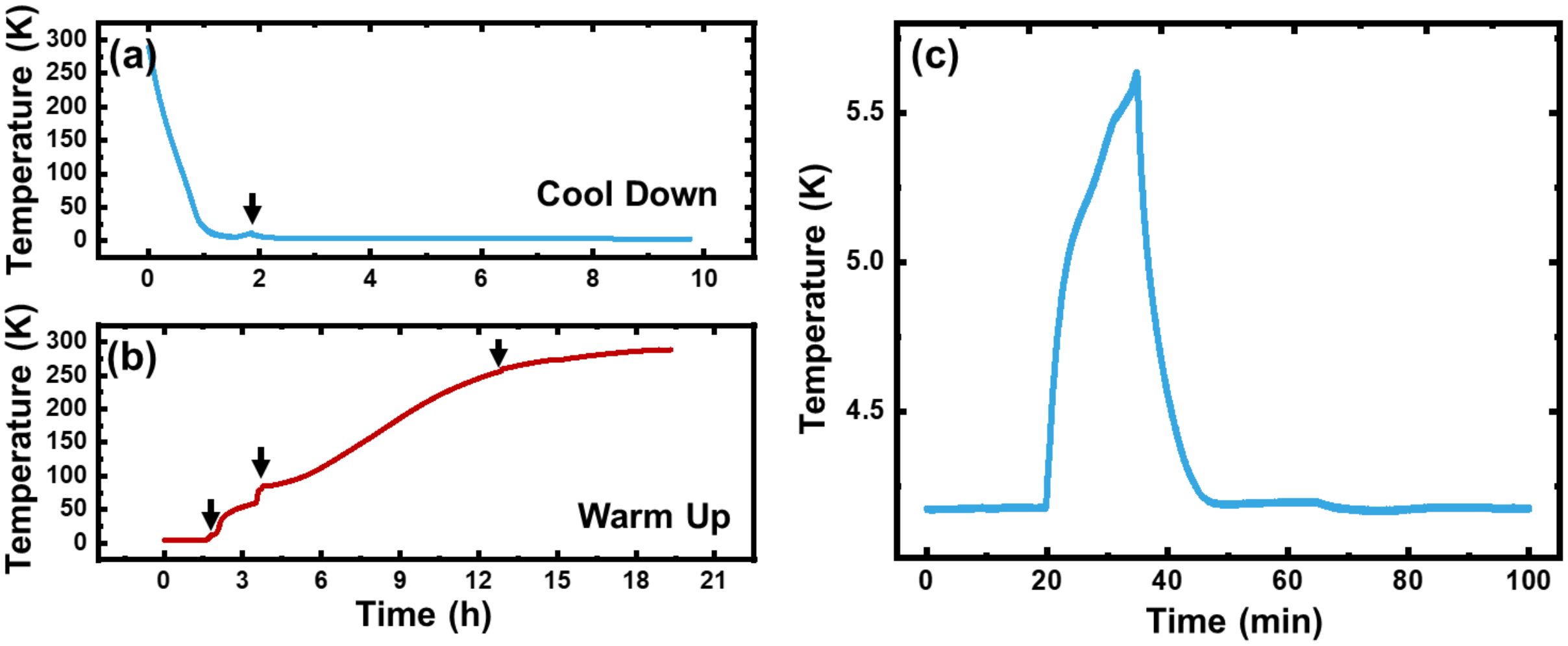

FIG. 3 Temperature vs. time curves recorded during (a) cooling down and (b) warming up the liquefier cold head, and (c) swapping the receiving and refilling dewars. The temperature discontinuities in the warming-up curve, indicated by the arrows, correspond to the phase change temperatures of hydrogen (14 K and 20 K), nitrogen (63 K and 77 K), and water (273 K). The small bump in the cooling-down curve (indicated by the arrow) marks the moment when all He vapor loads are connected to the recovery system, signaling the system's entering into its fully functional status.

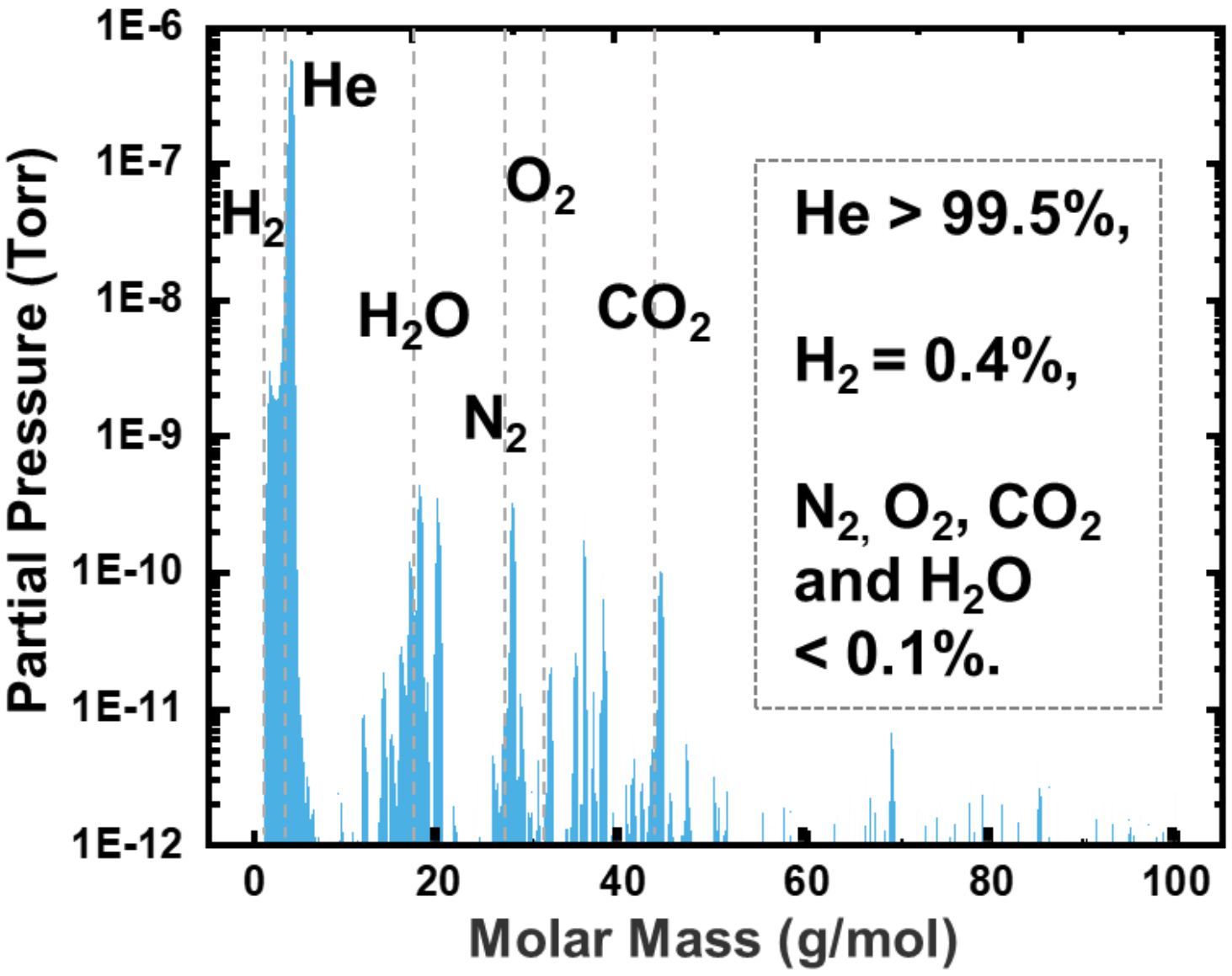

FIG. 4 Mass spectrum of the sampled He vapor compressed into the He boil-off storage tanks.

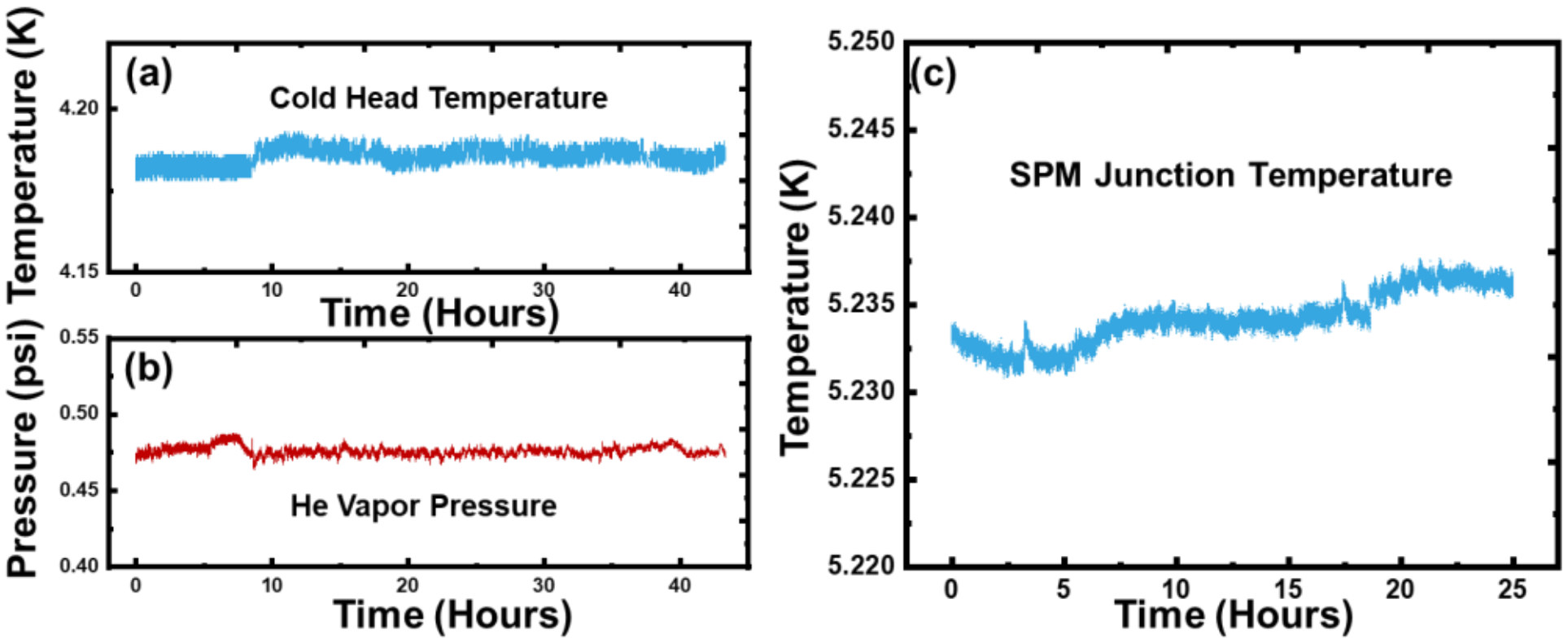

FIG. 5 Pressure and temperature stability of the He recovery system. (a) He vapor line pressure, (b) cold head temperature, and (c) SPM junction temperature during normal operation.

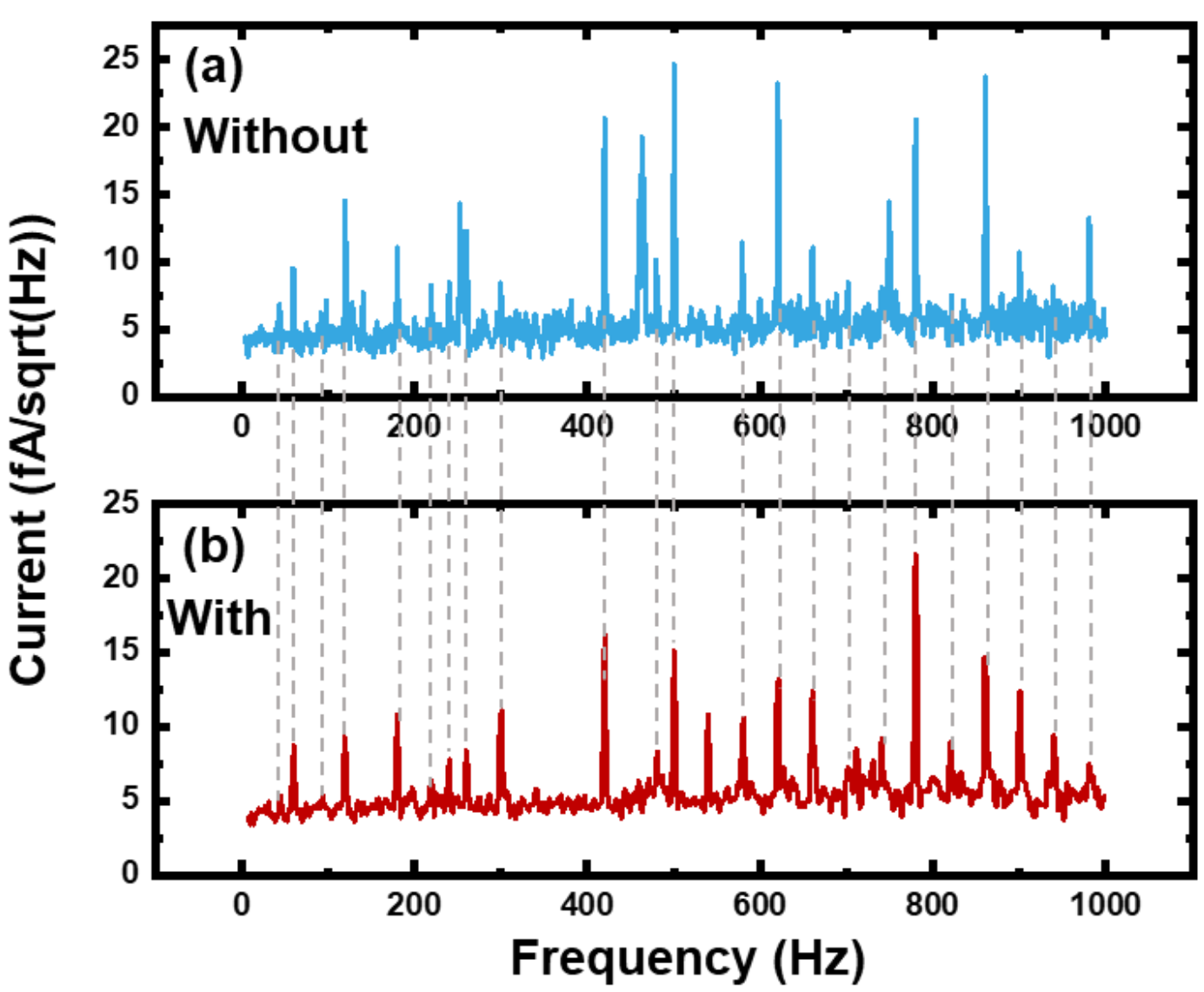

FIG. 6 FFT spectra of the tunneling current without (a) and with (b) being connected to the He recovery system. Each spectrum is an average of 10 scans.

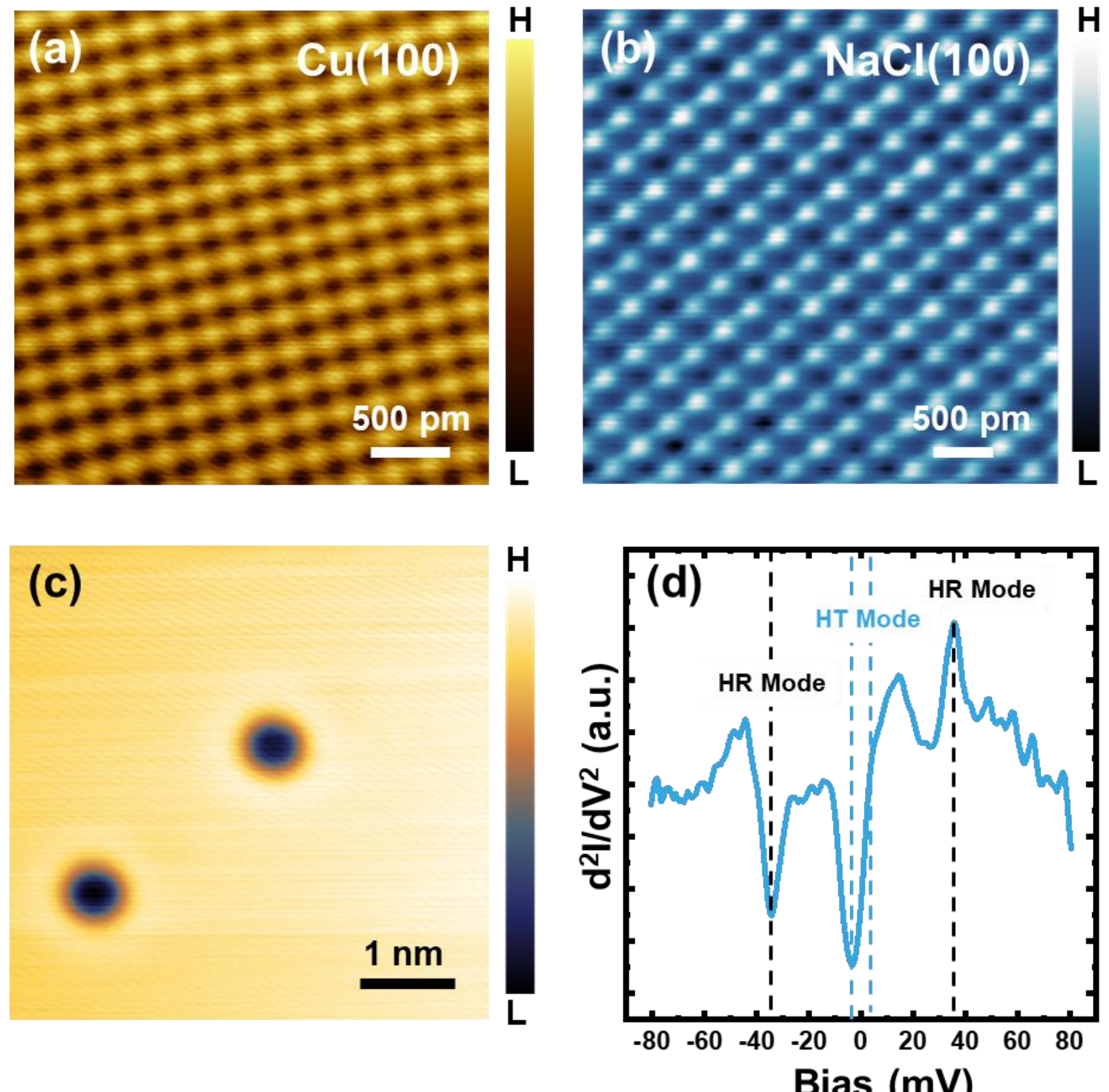

FIG. 7 Characterization of the SPM performance with the recovery system connected. STM topographic image of (a) Cu lattice at 60 mV, 500 pA set points, (b) NaCl lattice at -100 mV, 500 pA set points, and (c) two individual CO adsorbed on Cu(100) at -100 mV, 100 pA set points. (d) IETS of a single CO molecule. The energy of the HR and HT modes are marked by the dashed lines. The tunneling gap is set to 60 mV and 500 pA and the spectrum is an average of 20 scans.

## ACKNOWLEDGMENTS

This research was primarily supported by the National Science Foundation (NSF) under the grant number CHE-2303936. This research was partially supported by NSF through the UC San Diego Materials Research Science and Engineering Center (UCSD MRSEC) DMR-2011924. The authors also benefitted from the invaluable hardware supplies provided by and insightful discussions with Professor John Crowell at the University of California, San Diego.

## APPENDIX

### APPENDIX A: Connection details

#### A1. *The connection of LHe dewars to the He vapor line*

Both the receiving and refilling dewars, CryoFab CMSH LH100, are equipped with a liquid level meter and have been modified by adding a male Swagelok SSL Instrumentation Quick Connector (SS-QC8-S-8FT) to the ventilation port. During the normal operation time, the receiving dewar is attached to the He vapor line via a female connector (SS-QC8-B-810) close to the liquefier for recycling the boil-off gas (**Figure 8a**). The refilling dewar is connected to the He vapor line for recovery in a similar way but in the SPM room (**Figure 8b**). The self-lock mechanism of the quick connectors minimizes the leakage of He and the contaminations from the air when connecting/separating the dewars and the He vapor line. At the refilling dewar side, a manual angle valve is used to control the connection between the dewar and the He vapor line. Due to the frequent connections/disconnections of the refilling dewar, normally every 3 days for refilling the SPM, a Swagelok bellow valve is used to purge any possible contaminants after reconnecting the refilling dewar and when the manual angle valve is closed.

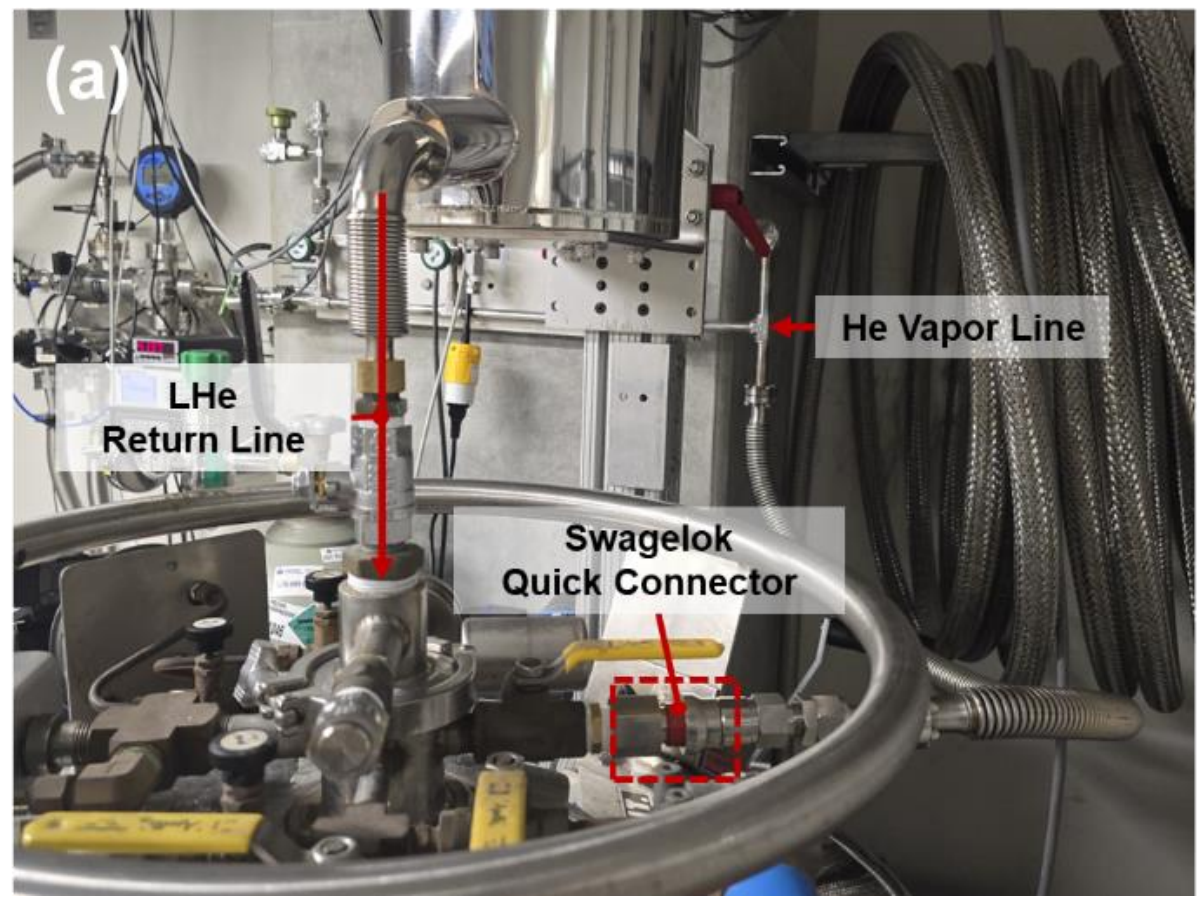

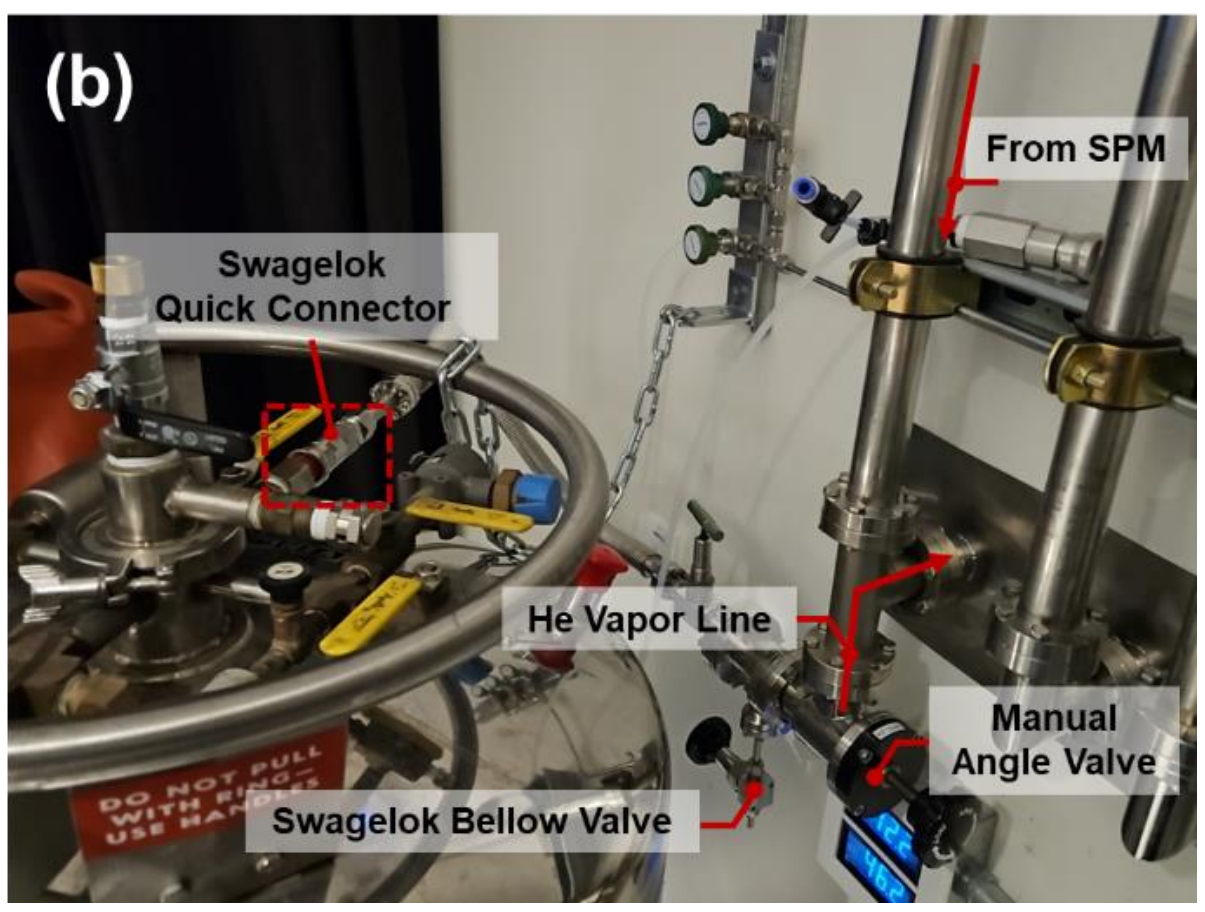

FIG. 8 The connection of LHe dewars to the He vapor line.

**A2.** ***The connection of SPM cryostat to the He vapor line***

Our SPM cryostat is terminated with a KF16 connector which is connected to the He vapor line via an SSL bellow for vibrational damping. We added a 3-psi safety valve (Generant vent relief valve, model VRV-500SS-V-3) close to the cryostat to prevent overpressure (**Figure 9**). Two manual angle valves are installed in the vicinity of the SPM cryostat. Valve 1 controls the connection between the He vapor line and the SPM cryostat while Valve 2 is used for venting only the SPM cryostat in some situations (**Appendix C2**). A type-K thermocouple is installed between the SPM cryostat and Valve 1 to monitor the temperature of rapid He boil-off during liquid transfer, which stably reads ~ -110 °C when there is liquid flowing from the refilling dewar to the SPM cryostat in our setup. Additionally, a household water pipe heating tape is wrapped on the He vapor line and powered during liquid transfer to alleviate the frostbite to the Viton seals on the manual angle valves by the cold He vapor.

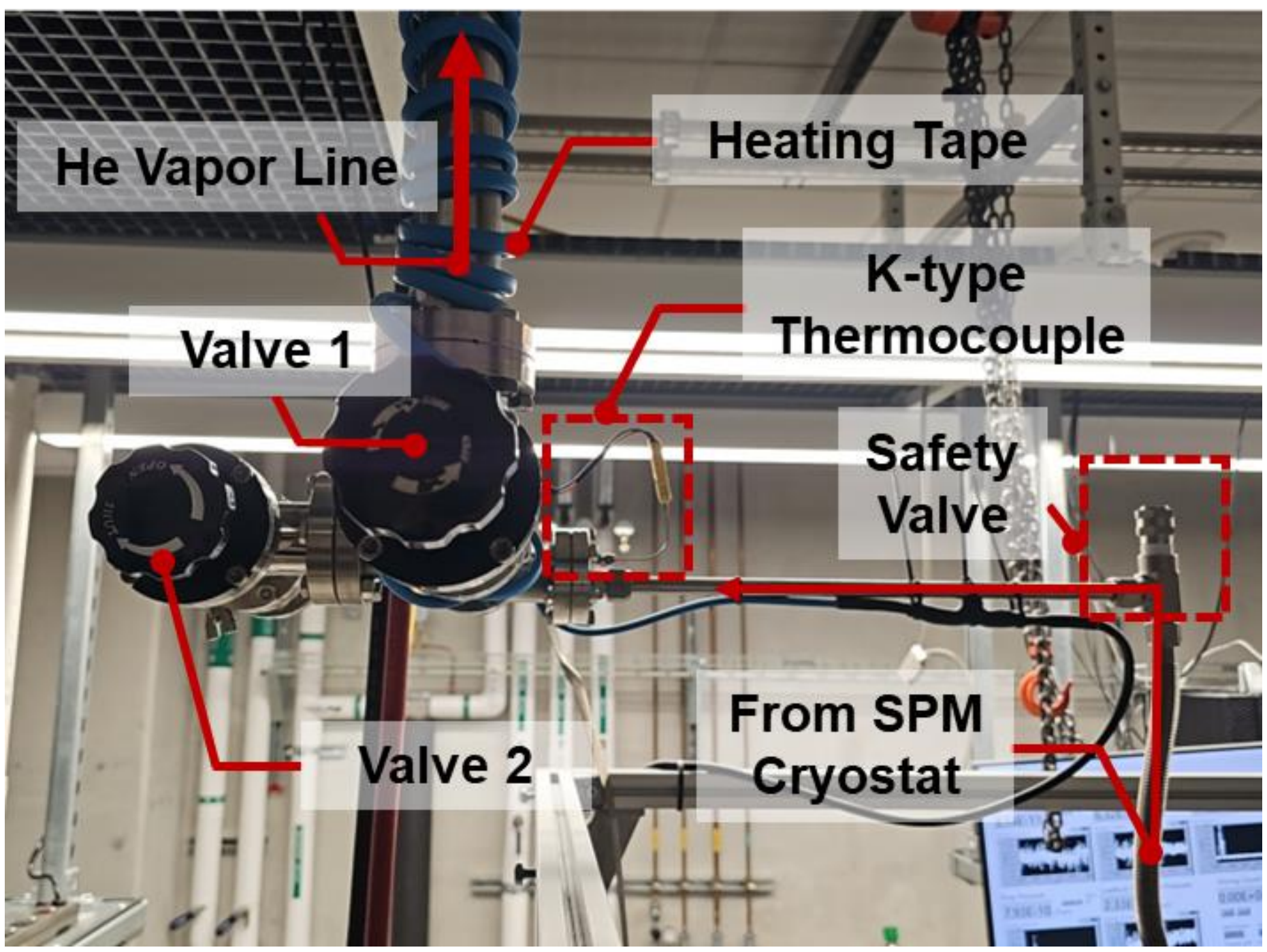

FIG. 9 Connection details close to the SPM cryostat.

**A3.** ***The connection of flow controller to the He vapor line***

The flow controller (MKS MFC GE50A) is terminated with VCR connectors at both sides. Its outlet is connected to a bellow valve on the He vapor line, while the inlet is linked to a Swagelok four-way cross, which is also connected to the UHP/industrial He cylinder at the right, the purifier at the left and a bellow valve for venting at the bottom (**Figure 10**).

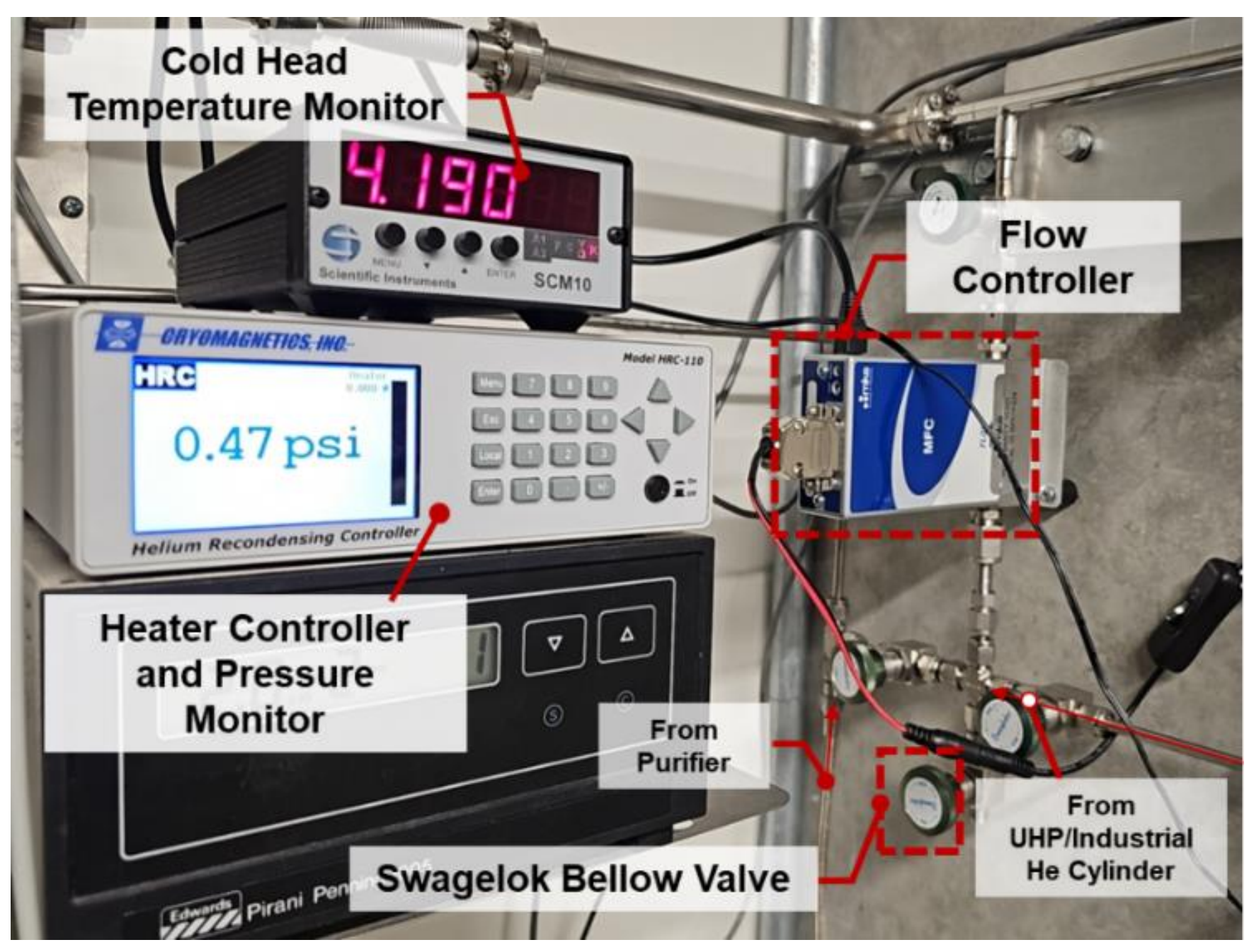

FIG. 10 Connection details of different He gas sources to the He vapor line via the flow controller.

**A4.** ***Air compressor and its connection to the He vapor line***

The compressor for compressing the excess He boil-off is slightly modified from an oil-free air compressor (California Air Tools 600040CAD). Specifically, to minimize both He leakage and air contamination, He-compatible sealant (Gasoila GE16) and Teflon tape have been applied to all NPT connections on the He boil-off storage tanks and the motor (**Figure 11**). The compressor came with two motors, but we found that one is powerful enough for our purpose. To increase the gas storage capacity, we attached two extension tanks, a 60-gallon, and a 10-gallon tank, to the original 60-gallon one and expanded it to 130 gallons (**Figure 11**). As discussed in the main text, the storage tanks can hold up to 4,677 L He gas (6.24 L LHe) at 125 psi and this capacity can be easily increased by incorporating more extension tanks. When it's close to 125 psi, the manual water exhaust valves at the bottom of the 60-gallon tanks can be used to release excess gas. The connection between the air compressor motor and the He vapor line is controlled by a normally closed pneumatic valve (MDC AV-150M-P) (**Figure 11**). We also modified the power box of the air compressor to control it with a solid-state relay. The on and off state of both compressor motor and pneumatic valve are triggered by our LabVIEW program, as detailed in **Appendix B**.

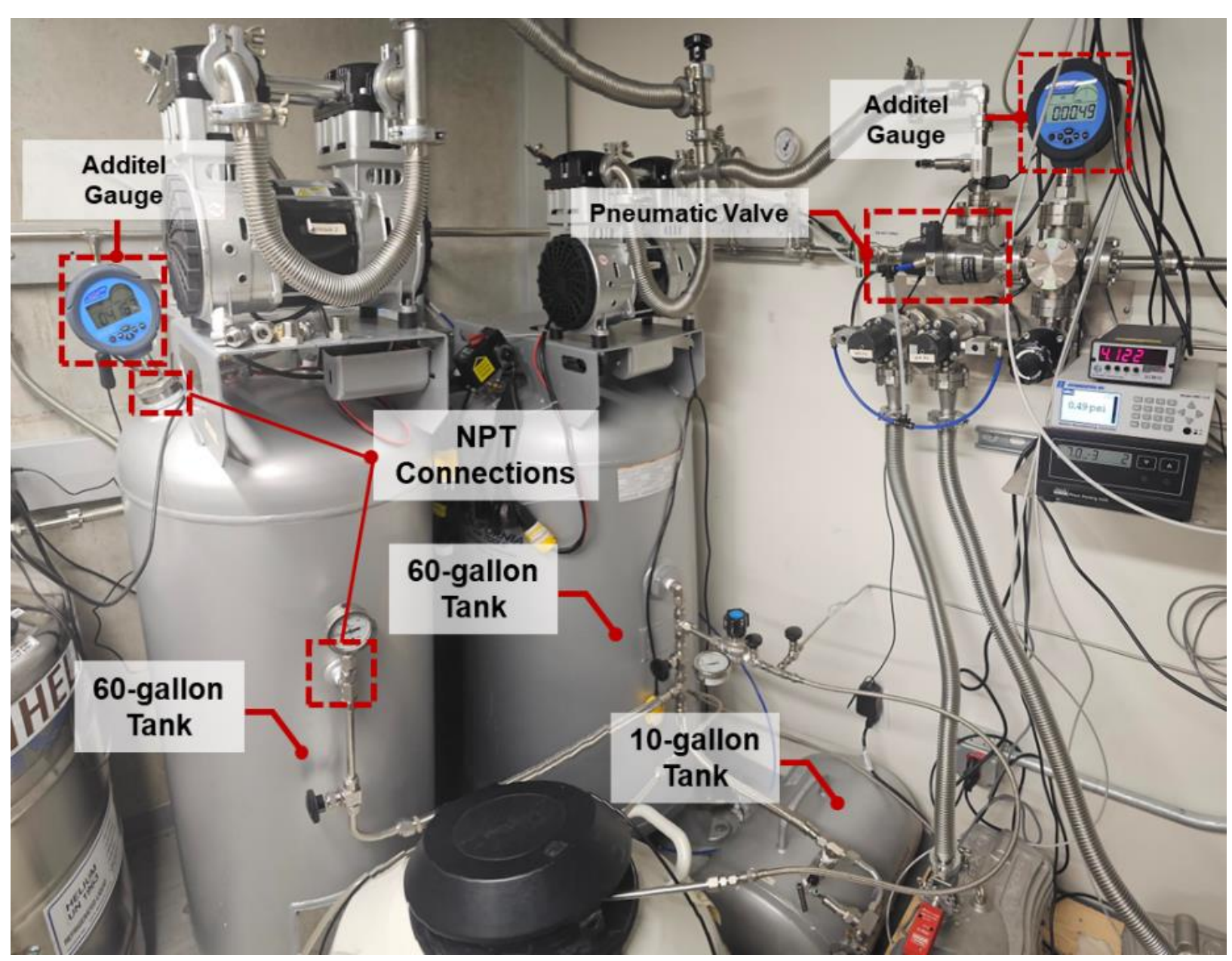

FIG. 11 Picture of the air compressor and its connection to the He vapor line.

**APPENDIX B: LabVIEW program for stabilizing He vapor pressure and controlling the air compressor**

We have three measures to stabilize the He vapor pressure. The inverter and heater included in the liquefier package offer two ways to target the He pressure to a preset value, as discussed in the main text. They both use the reading from a pressure transducer (SSI Technologies P51) installed close to the liquefier inlet as feedback. In addition, we wrote a LabVIEW program to help maintain the He pressure. It reads the He vapor line and storage tank pressure from pressure gauges (Additel 681), the heater power and cold head pressure from He recondensor controller (HRC-110), the cold head temperature from the single channel temperature monitor (SCM-10) and the status of the cold head compressor package (CPA286I) by constantly querying the data via serial connections (RS232). The program communicates with the flow controller via ethernet connection (TCP-Modbus) and controls its opening according to the pressure readings from both Additel pressure gauges. Additionally, the LabVIEW program can switch on/off the pneumatic valve and air compressor by sending analog signals to the solid-state relay with a National Instruments data acquisition device (NI BNC-2120 DAQ). Specifically, when the He vapor pressure exceeds a “pressure upper limit”, e.g., 1.6 psi, the LabVIEW program will first turn the compressor on and then open the pneumatic valve to compress the excess He boil-off. When the He vapor pressure drops below a “pressure lower limit”, e.g., 0.4 psi, the LabVIEW program will first close the pneumatic valve and then turn the compressor off to avoid venting the He vapor line. **Figure 12** shows a screenshot of the LabVIEW program with the flow control module and the air compressor control module highlighted.

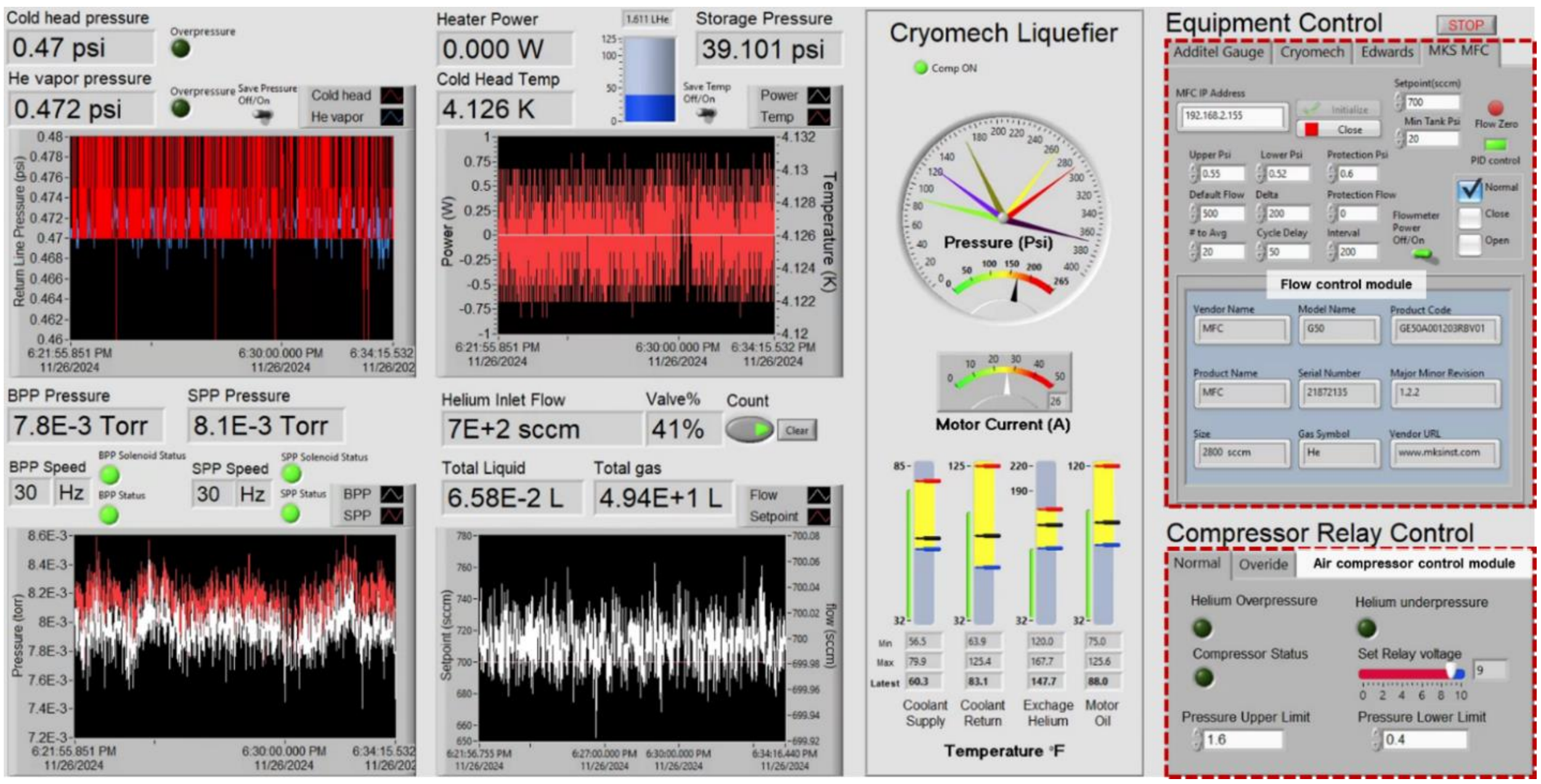

FIG. 12 Screenshot of our LabVIEW interface.

Shown in **Figure 13** are the schematics of compressing the He boil-off into the storage tanks. Initially, the pneumatic valve is closed, the compressor motor is vented to the air and the He boil-off tank is sealed by the check valve on the top (**Figure 13a**). Once the compressor motor turns on, the solenoid valve closes and then the pneumatic valve opens to let the He boil-off in. To remove the residual air, we keep the manual ball valve open in the first few seconds (**Figure 13b**). Then, we close it to redirect the He flow into the boil-off storage tank (**Figure 13c**). When the He vapor pressure drops below the "pressure lower limit", the pneumatic shuts down, and then the compressor motor stops while the solenoid valve opens to vent this section. Afterward, we manually open the ball valve for the next compression (**Figure 13a**).

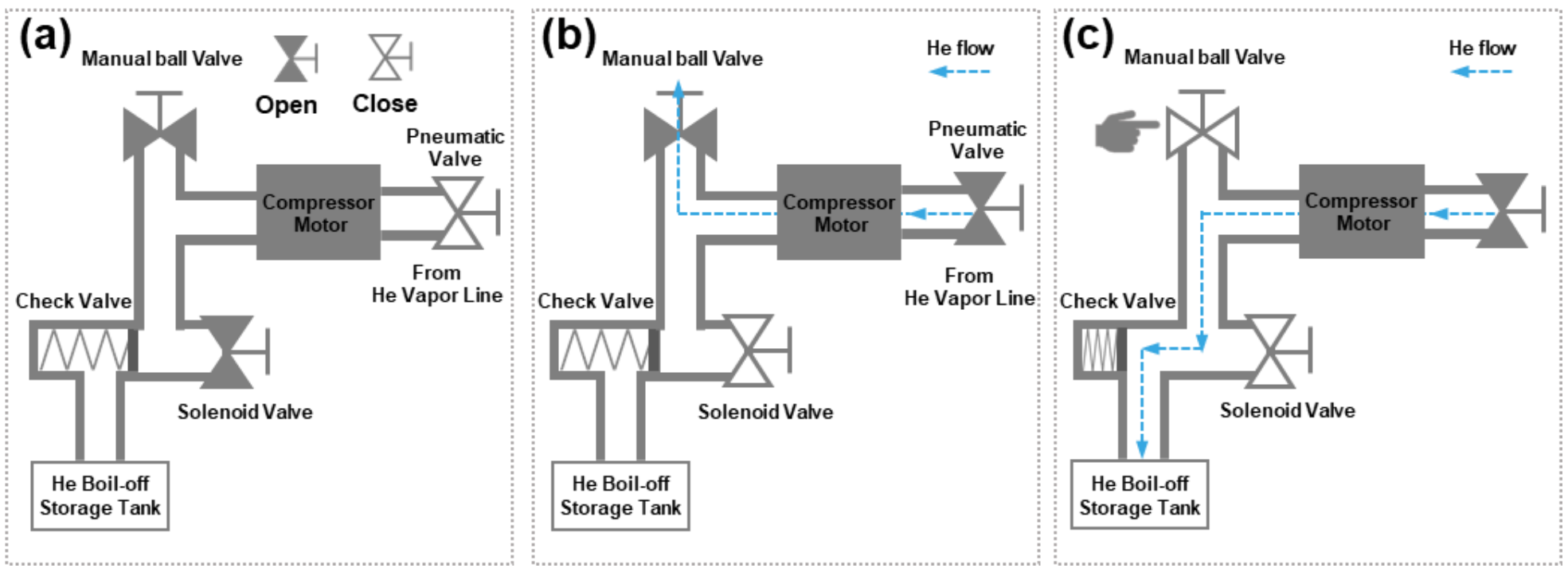

FIG. 13 Schematics of compressing He boil-off into the storage tanks.

**APPENDIX C: Operation details**

**C1.** ***Initial cooldown of the receiving dewar***

During the initial cooldown of the receiving dewar as discussed in the main text, recycling the large amount of He boil-off is desirable to increase the recovery rate. To achieve this, we first cooled down the receiving dewar with $LN_2$, then purged with dry nitrogen, followed by UHP He gas purging according to the standard procedures. Afterward, we transferred LHe from the refilling dewar, which had several tens of liters of LHe, to the receiving dewar for further cooldown. After a short while, we connected the receiving dewar to the He vapor line to have the air compressor compress the He boil-off until the storage tanks were full. This cooldown process stopped when the level meter started reading the LHe level in the receiving dewar, which was then attached to the He vapor line to help cool down the liquefier cold head, as described in the main text.

**C2.** ***Initially cooling down and refiling SPM cryostat***

As shown in **Figure 9**, normally, Valve 1 is open while Valve 2 is closed to direct the boil-off from the SPM cryostat to the liquefier. However, when initially cooling down or refilling the SPM cryostat, Valve 1 needs to first be closed to avoid venting the whole He vapor line and Valve 2 has to be open to release the pressure in the cryostat. Once the LHe transfer line is inserted into the cryostat, one can shut down Valve 2 and then open Valve 1 to redirect the momentary He boil-off to the He vapor line for compression by the air compressor. Once the LHe transfer finishes, Valve 1 should be closed and Valve 2 should be open for pulling the transfer line out from the cryostat. Afterward, the SPM cryostat is connected to the He vapor line again for recycling by closing Valve 2 and opening Valve 1.

**C3.** ***Regenerating the purifier***

The SSL coil immersed in the $LN_2$ can be clogged by the contaminants in the He from the boil-off storage tank over time. In this case, there won't be enough He from the coil outlet; thus, the flow controller will be fully open to try to meet the preset flow. The coil can be unclogged in two ways. An easy and fast way is to first shut off the flow controller and then lift the top two loops of the coil up from the surface of $LN_2$ while blowing them with a hair dryer until they are defrosted (**Figure 14**). Any potential contaminants escaping from the coil and heading towards the flow controller during this process will be purged via the Swagelok bellow valve right below the flow controller (**Figure 10**) before redirecting the purified He into the recycling line. The cryogenic freezer (MVE XC47/11-6) needs to be filled with $LN_2$ again afterwards. If this method doesn't work as effectively any more due to the accumulated frozen contaminants in the lower loops of the coil, we simply replace the dirty coil with a clean one. This is made reproducible by using Swagelok connections. Once connected, one can purge the new coil with the He from the boil-off storage tank before immersing it into the $LN_2$. Before feeding to the He vapor line, the contaminated He between the coil outlet and flow controller has to be purged with the purified He out from the coil.

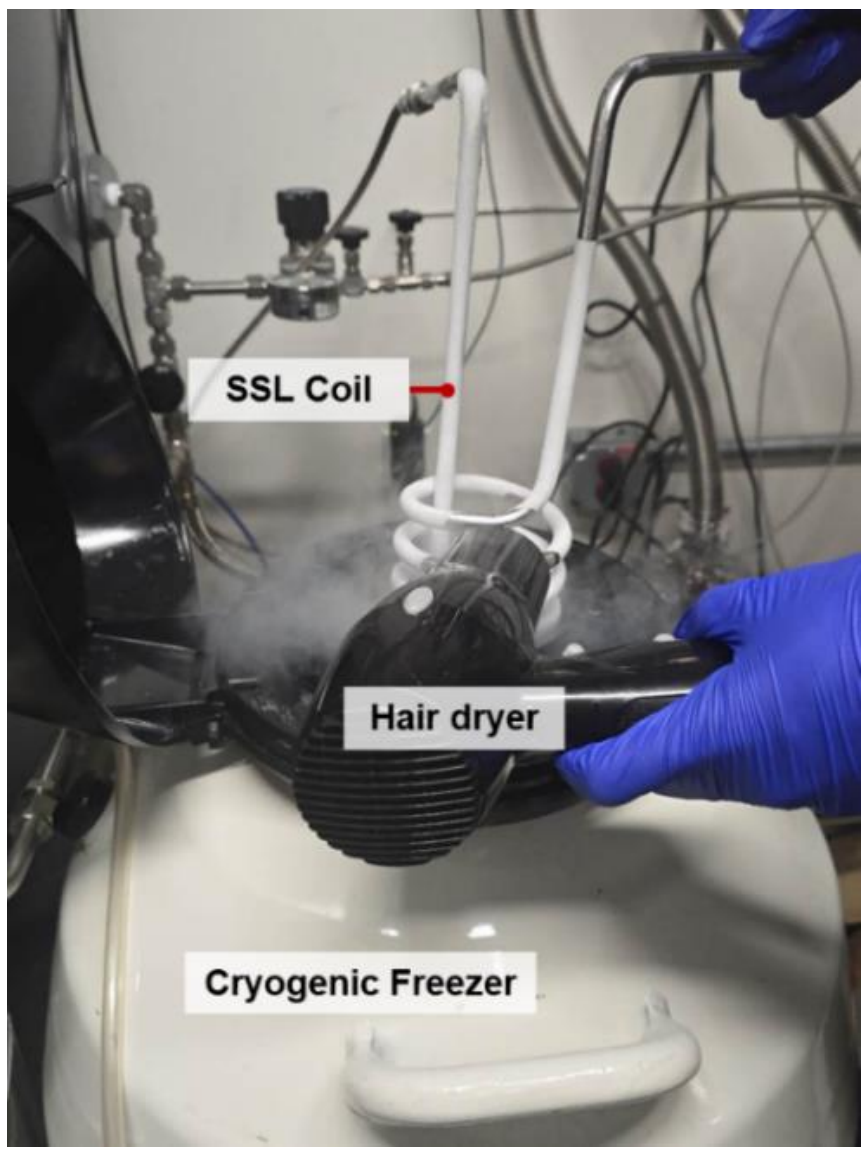

FIG. 14 Image of unclogging the coil.

**C4.** ***Swapping the receiving dewar and the refilling dewar***

Once the receiving dewar is close to full, we swap it with the refilling dewar, which is about to be empty. In this case, we first isolate the SPM cryostat and disconnect the refilling dewar from the He vapor line. Usually, the inline He pressure will quickly drop, and the cold head compressor will run at minimal power due to the decreased He boil-off load at this stage. Then we quickly pull the return line out from the receiving dewar without turning off the cold head compressor and keep a 1500 sccm UHP He flows through the return line via the flow controller. When the return line is fully out from the receiving dewar, we blow its outlet with a heat gun to prevent frosting while quickly exchanging the receiving dewar and the refilling dewar. Afterward, we insert the return line into the new receiving dewar (the previous refilling dewar) and attach the new refilling dewar (the previous receiving dewar) to the He vapor line in the SPM room.

**C5.** ***Decontaminating the cold head***

When the cold head performance is compromised by contaminants, we either warm it up to 80 K or room temperature. In this scenario, we first isolate the cold head from both the SPM cryostat and refilling dewar with the manual angle valves (**Figure 8b** and **9**). Then, we turn the cold head compressor off and pull the return line out from the receiving dewar until it's above the ball valve at the refilling port of receiving dewar. Sequentially, we close the ball valves at both the refilling port and vent port of the receiving dewar and vent it via a 0.5 psi safety valve (Generant Vent Relief Valve, VRV-250B-V-.5). At this stage, the return line outlet is still isolated from the air by sealing it with the O ring at the refilling port of the receiving dewar. When the cold head warms up, the He vapor line will be pressurized and we vent it through a 0.5 psi safety valve (Generant Vent Relief Valve, VRV-250B-V-.5) installed close to the liquefier (**Figure 15**). We usually turn on the heater to shorten the warming-up time to 80 K. If only warming up to 80 K, we then fully pull the return line out from the receiving dewar refilling port, purge the cold head with UHP He, insert the return line back into the receiving dewar and turn on the cold head compressor. If warming up to room temperature, we fully pull the return line out from the receiving dewar at 80 K and purge the cold head with dry nitrogen via a port close to the liquefier (**Figure 15**) to accelerate the warming up process. Once the cold head is at room temperature, we purge it with UHP He, insert the return line back into the receiving dewar and turn on the cold head compressor, which is similar to initial cooldown of the cold head.

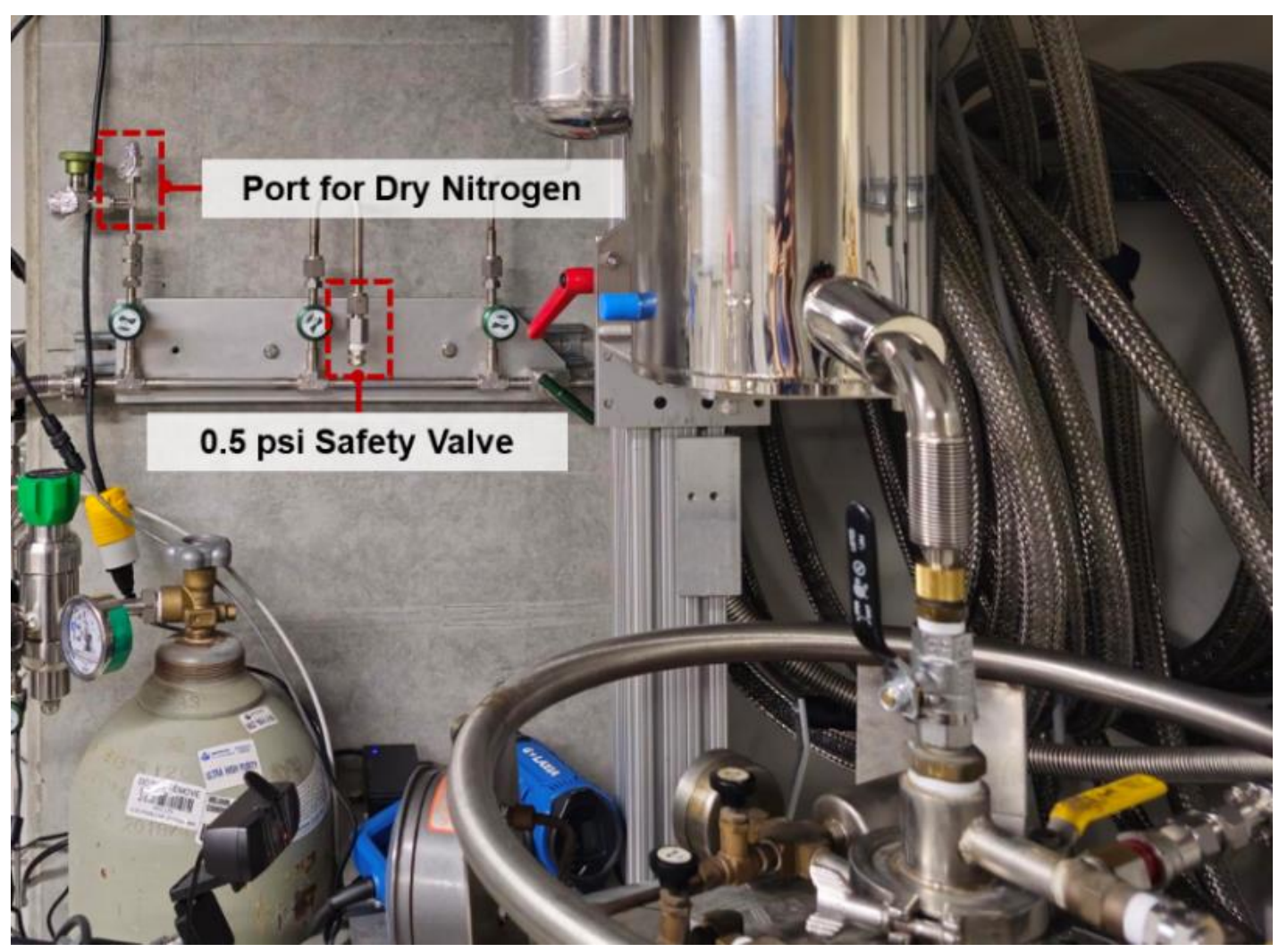

FIG. 15 Picture of the connections close to the liquefier.

**C6. *Replenishing LHe***

LHe in our recovery system can be replenished by directly attaching a commercial LHe dewar to the He vapor line (**Figure 16**) or liquefying the UHP/industrial He from the high-pressure cylinders. Given the low consumption rate of our system, ~0.27 L/day, the latter is preferred because otherwise the receiving dewar will get full sooner and we need to swap the dewars more frequently, which potentially introduce more contaminants to the recovery system and increase the consumption rate.

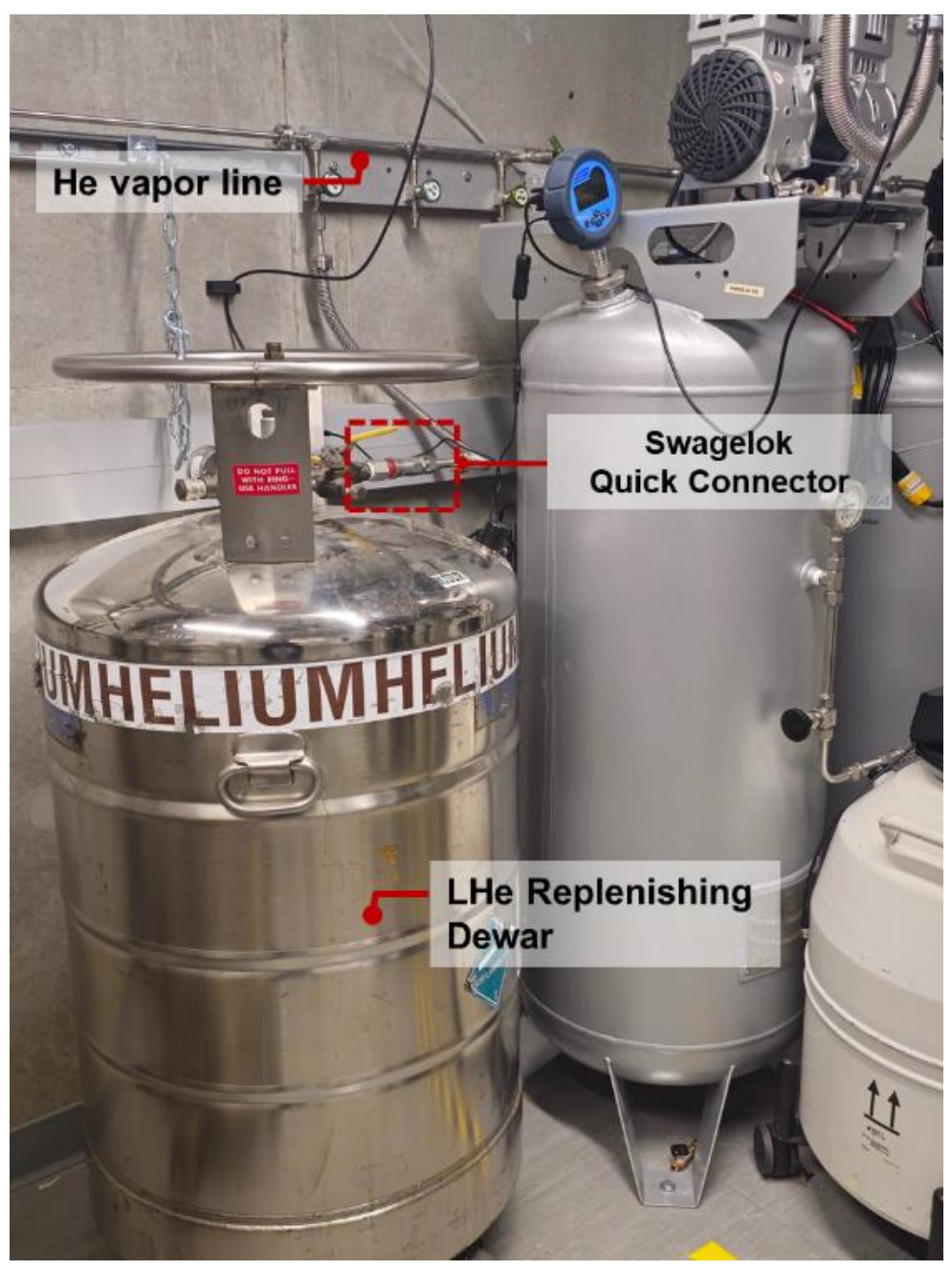

FIG. 16 Picture of the connection of the LHe replenishing dewar to the He vapor line.

## AUTHOR DECLARATIONS

### Conflict of Interest

The authors have no conflicts to disclose.

## DATA AVAILABILITY

The data that support the findings of this study are available from the corresponding authors upon reasonable request.